\documentclass[pra,twocolumn,preprintnumbers,amsmath,amssymb]{revtex4}

\usepackage{graphicx}
\usepackage{epsfig}
\usepackage{dcolumn}
\usepackage{bm}

\begin{document}

\title{Pseudo-digital quantum bits}

\author{Mark Friesen}
\email{friesen@cae.wisc.edu}
\author{Robert Joynt}
\author{M.~A.~Eriksson}
\email{maeriksson@facstaff.wisc.edu}
\affiliation{Department of Physics, University of Wisconsin, Madison, 
Wisconsin 53706, USA }

\begin{abstract}
Quantum computers are analog devices; thus they are highly susceptible to accumulative errors arising from classical control electronics.  Fast operation--as necessitated by decoherence--makes gating errors very likely.  In most current designs for scalable quantum computers it is not possible to satisfy both the requirements of low decoherence errors and low gating errors.  Here we introduce a {\it hardware}-based technique for pseudo-digital gate operation.  We perform self-consistent simulations of semiconductor quantum dots, finding that pseudo-digital techniques reduce operational error rates by more than two orders of magnitude, thus facilitating fast operation.
\end{abstract}


\maketitle

Quantum computation would not be feasible without quantum error correction.\cite{preskill98}  However fault-tolerant codes can succeed only when error rates fall below a certain threshold.\cite{steane02}  Error management therefore involves {\it prevention} as well as correction.  In this letter we focus on the prevention of quantum gate errors.  There are two possible approaches: perform the gates slowly (and thus accurately), or quickly (with associated bandwidth noise) using hardware that is relatively noise-tolerant.  NMR quantum computers have succeeded in performing some simple quantum computations, by taking the slow approach.\cite{vandersypen01}  However, promising applications like the factorization of large numbers\cite{shor94} involve many quantum gate operations.  Indeed, more than $10^9$ Toffoli gates are required for factorizing a modest 130-digit number, using fault-tolerant codes.\cite{preskill98}  Such computations may take years to complete at NMR speeds.  Further, NMR qubits cannot be scaled up with current technologies.  It is therefore important to develop robust gating techniques for faster, scalable technologies, including spin qubits in semiconductors\cite{loss98,kane98} and superconducting qubits.\cite{makhlin01}

We consider here a specific architecture:  silicon-germanium quantum dots containing single electrons.\cite{loss98,hu00,burkard99,levy01,friesen02}  The electrons are confined vertically by a quantum well heterostructure and laterally by repulsive electrostatic interactions with lithographically patterned top-gates.  Voltage pulses lower the potential barrier between pairs of quantum dots, moving the electrons towards each other and ``turning on" their exchange coupling $J$.\cite{loss98,hu00,burkard99}  Details of the device are given in Ref.~\onlinecite{friesen02}.  

The analog voltages that control couplings between the spin qubits cannot be produced perfectly, resulting in errors.\cite{friesen02}  We consider relative uncertainties in the voltage height $\Delta V/V$, arising from classical control electronics.  Such control errors are unavoidable--in practice, they increase with the operating speed.  The voltage $V$ controls two-qubit gate operations by ``turning on" the qubit coupling $J$.  If the dependence of $J$ on $V$ is strong, then even small voltage uncertainties lead to large coupling errors.  

The slope $\partial J/\partial V$ determines the susceptibility of a quantum dot device to voltage errors.  The dimensionless susceptibility is given by 
\begin{equation}
\Omega = \left| \frac{V}{J} \frac{\partial J}{\partial V} \right| ,\label{eq:omega}
\end{equation}
where $V$ is the pulse height.
$\Omega$ is useful because it converts fractional voltage uncertainties into fractional gating errors:  $\Delta J/J = \Omega \Delta V/V$.  For virtually all qubit proposals, $\Omega$ is greater than zero, and it may be quite large for particular designs.  Corresponding error rates are large, causing error correction techniques to fail.   The requirement for fault tolerant computation is \cite{preskill98,note1}
\begin{equation}
\Omega \frac{\Delta V}{V} < 10^{-4} .\label{eq:faulttol}
\end{equation}

The solid curve in Fig.~1 shows the dependence of exchange coupling $J$ on voltage $V$, as computed for the conventionally-gated quantum dot of Ref.~\onlinecite{friesen02}.  From Eq.~(\ref{eq:omega}) we obtain $\Omega =5$-10.  The reason $\Omega$ is large in this architecture (and other spin qubit architectures) is that the qubit coupling depends exponentially on the spacing between the two electrons, due to the exponential shape of the wavefunction tails.  The strong dependence of $J(V)$ translates into a large value of $\Omega$.  Consequently, small uncertainties in the gate voltage produce substantial coupling errors.  Assuming $\Omega =5$-10, at the GHz operation speeds needed to beat decoherence $\Delta V/V$ is so large that Eq.~(\ref{eq:faulttol}) cannot be satisfied.\cite{friesen02}
\begin{figure}
\centerline{\epsfxsize=2.5in \epsfbox{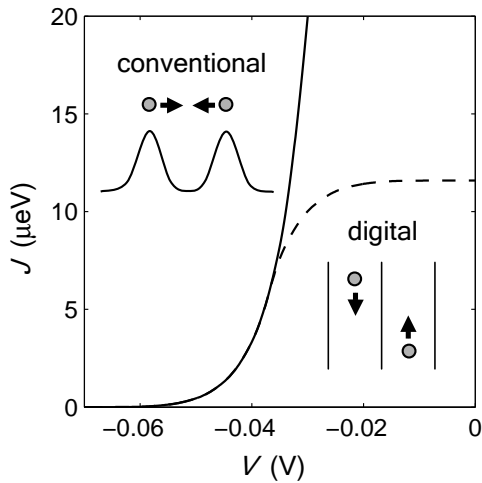}}
\caption{
Comparison of a conventional voltage-coupling function (exponential curve) and a hypothetical digital response function (flattop curve).  The conventional response was computed in Ref.~\onlinecite{friesen02}, where the voltage $V$ controlled the barrier height, and thus the overlap, between two quantum dots.  The resulting exchange coupling $J$ was exponential in the qubit separation, due to the change in overlap between the electron wavefunctions (upper inset).  A flattop response that is robust against gating errors can be obtained by sliding qubit electrons transversely in channels (lower inset).}
\end{figure}

A truly digital dependence of $J$ on $V$ (Fig.~1, dashed curve) would give $\Omega \simeq 0$, facilitating fault-tolerant computation.  Such a response is obviously desirable, but unfeasible in an analog device.  Here we propose a {\it pseudo-digital} technique, in which the electrons are confined to separate channels (Fig.~1, lower inset), such that they move past each other instead of directly towards each other.\cite{leuenberger01}  There is a point of minimum separation where the coupling $J$ is a maximum and its dependence on $V$ is negligible.  In other words, $J(V)$ becomes flat-topped.  It is not immediately obvious that such behavior can be realized.  In the following, we propose a specific pseudo-digital design for quantum dot qubits, and demonstrate its feasibility through detailed simulations of a realistic silicon-germanium device.

The main challenge for designing channel qubits is to provide adequate travel for the sliding electrons.  Ideally, when the qubit coupling is turned ``off," the electrons should be well separated.  In the ``on" configuration, the coupling should be large enough to enable fast operation.  To meet this challenge we have developed a bistable quantum dot design (Fig.~2).  Using opposing plunger gates, an electron is squeezed into either side of a dot.  The flattop ``on" condition for $J(V)$ corresponds to the side-by-side configuration of electrons in neighboring dots.

To determine the feasibility of the channel qubit architecture, we model the double quantum dot device of Fig.~2(b), containing one electron per dot.  The heterostructure is the same as in Ref.~\onlinecite{friesen02}.  Precise Poisson and Schr\"{o}dinger equation simulations are performed using finite element techniques.  A basis set of single-electron wavefunctions is computed  for each qubit in the Hartree-Fock approximation.  A two-electron basis set is obtained in the configuration interaction approach, and exact diagonalization is performed on the corresponding Hamiltonion matrix.  Image potentials arising from the various gates, due to the qubit electrons, are computed using Green's functions techniques.  The computations are then iterated until self-consistency is achieved.  We do not consider impurities here, or other inevitable imperfections in the crystal lattice.  However we note that disorder cannot invalidate the basis of pseudo-digital control--the existence of a minimum in the separation between the two qubits. 
\begin{figure}
\centerline{\epsfxsize=3.35in \epsfbox{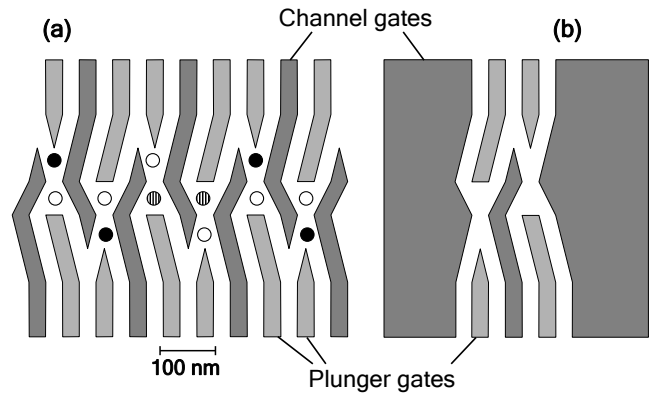}}
\caption{
Top-gate design for implementing a channel architecture using bistable quantum dots:  (a) scaled-up design with many qubits, (b) two-qubit design, as simulated in this paper.  Opposing plunger gates (light gray) push electrons into the top or bottom half of a given dot.  Voltages on the channel gates (dark gray) remain constant and highly accurate, controlling the maximum electron exchange coupling.  Coupling is ``off" for up-down configurations (solid circles), and  ``on" for side-by-side configurations (striped circles).  Plunger gates can be operated fast, because pseudo-digital techniques make the coupling $J$ insensitive to plunger gate voltage uncertainties.}
\end{figure}

Fig.~3 shows the simulation results.  As the plungers force the electrons into their side-by-side configuration, the exchange coupling grows, exponentially at first.  The coupling reaches a peak as the qubits slide past one another; this is the flattop condition we exploit for pseudo-digital control.  The bistable dot design clearly enables the desired range of electronic motion, from an ``off" state, where the qubit coupling is exponentially small, to an ``on" state, where $J(V)$ is flat-topped.  

For the device studied here, the exchange coupling achieves a maximum value of $0.4~{\rm \mu}$eV, corresponding to gate time of 5~ns for the two-qubit SWAP.  Faster operation can be achieved with appropriate modifications to the qubit design.  In the present work, we have assumed minimum feature sizes of 30~nm for lithographic patterning, giving a qubit separation of 90~nm for the ``on" configuration.  Since $J$ decreases exponentially with qubit separation, reducing the feature dimensions will increase both $J$ and the gating speed.

One of the key features of the qubits in Fig.~2 is that the plunger gates and the channel gates perform very different roles.  The channel gates regulate the maximum value of $J$.  Since the channel gate voltages are held constant, this can be accomplished with arbitrary precision.  Switching $J$ on and off is the function of the plunger gates, as controlled by classical control electronics.  Ideally, plunger gate voltages will be varied quickly, near GHz frequencies.  This speed makes them subject to the large uncertainties $|\Delta V/V | \sim 0.01$ associated with high-speed electronics.\cite{friesen02,note2}  The key advantage to the shape of $J(V)$ in Fig.~3 is that first order uncertainties in $V$ translate to second order errors in $J$.  To estimate the errors in $J$, we calculate the RMS deviation of $J(V)$ from its mean, assuming a uniform distribution of gate voltages in the range $\pm \Delta V$.  This gives $\Delta J/J \simeq 5\times 10^{-4}$, or $\Omega \simeq 0.05$.  Thus, the channel qubit design decreases coupling errors by two orders of magnitude compared to conventional gating architectures.  This improvement is accomplished in spite of the fact that the $J(V)$ curve in Fig.~3 is only ``flat" at a single point.

With these results, Eq.~(\ref{eq:faulttol}) suggests that coupling errors $\Delta J$ will only need to be reduced by a factor of 5 to meet the $10^{-4}$ threshold level for fault tolerance.  If $10^{-3}$ is more appropriate, then the present scheme will be adequate.\cite{steane02}  This can be accomplished by improvements in existing pulse generators, emphasizing the dependence of scalable quantum computers on cutting-edge classical technology.  However, if enhancements in control electronics are not possible or desirable, $\Omega$ may be further reduced by optimizing quantum dot designs.  For example, top-gate patterns can be developed that cause qubit confinement potentials to become steeper.  Small voltage fluctuations will then have less of an impact on the electron positions, resulting in smaller fluctuations in $J$.  

In general, both single and two qubit gates are required for universal quantum computation.  So far, we have only demonstrated how to apply pseudo-digital techniques to two-qubit operations.  Fortunately, two qubit operations are known to be universal for {\it coded} spin qubits.\cite{bacon00,divincenzo00}  Thus, universal quantum gate operations can be controlled pseudo-digitally.
\begin{figure}
\centerline{\epsfxsize=2.5in \epsfbox{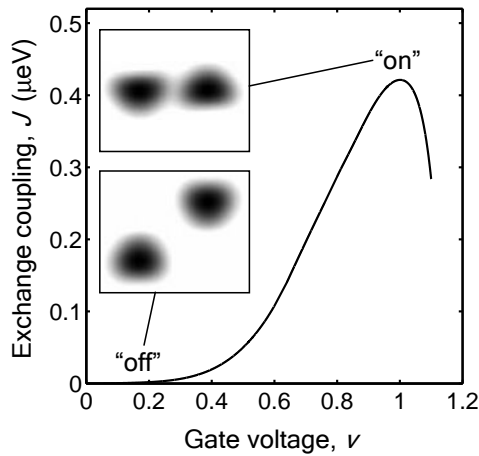}}
\caption{
Flattop coupling function $J$ vs. normalized gate voltage $v$, as computed for the gate architecture of Fig.~2b and the semiconductor heterostructure of Ref.~\onlinecite{friesen02}.  Insets show resulting charge densities at the extrema of $J$:  coupling ``off" ($v = 0$) and coupling ``on" ($v = 1$).\cite{note3}}
\end{figure} 

In this paper, we have shown how to construct qubits that enable fast gating by pseudo-digital techniques.  For any quantum computer, gate operations must be performed $10^4$ times faster than the decoherence rate.  However, unless special care is taken, fast operation results in large gate errors.  In general, an operational ``sweet spot" must exist where two fundamental constraints are satisfied:  the quantum gates are fast enough to beat decoherence and slow enough to be robust.  Here we have developed a hardware design for quantum dot spin qubits that helps meet these criteria by increasing the speed at which gating remains robust.  The concept is general and can be extended to many physical systems.\cite{unpub}

\begin{acknowledgments}
We have benefited from many discussions within the solid-state quantum computing group at the University of Wisconsin.  Our work was supported by the U.S. Army Research Office through the ARDA program, and the National Science Foundation through the MRSEC and QuBIC programs.
\end{acknowledgments}



\begin{thebibliography}{99}
\bibitem{note1}
The error threshold of $10^{-4}$ quoted here is an estimate.  More accurate figures can be derived with specific knowledge of the quantum computing hardware; see Ref.~\onlinecite{steane}.
\bibitem{note3}
The plunger gate voltages depend linearly on $v$.  Opposing pairs of plunger gates are used to push/pull the electron wavefunction.
\end{thebibliography}
\end{document}